%% file: mnras_template.tex
\documentclass[fleqn,usenatbib]{mnras}
\usepackage{newtxtext,newtxmath}
\usepackage[T1]{fontenc}
\usepackage{ae,aecompl,nicefrac}
\usepackage{graphicx}	
\usepackage{amsmath}	
\usepackage{amssymb}	
\usepackage{enumitem}
\usepackage{flushend}

\include{mnras_jdf}

\title[Embedded operator splitting methods]{Embedded operator splitting methods for perturbed systems}

\author[Hanno Rein]{
Hanno Rein$^{1,2}$\\
$^1$ Department of Physical and Environmental Sciences, University of Toronto at Scarborough, Toronto, Ontario M1C 1A4, Canada\\
$^2$ Department of Astronomy and Astrophysics, University of Toronto, Toronto, Ontario, M5S 3H4, Canada\\
}

\date{Submitted: 6 December 2019. Accepted: 22 January 2020.}

\pubyear{2019}

\begin{document}
\label{firstpage}
\pagerange{\pageref{firstpage}--\pageref{lastpage}}
\maketitle

\begin{abstract}
It is common in classical mechanics to encounter systems whose Hamiltonian $H$ is the sum of an often exactly integrable Hamiltonian $H_0$ and a small perturbation $\epsilon H_1$ with $\epsilon\ll1$.
Such near-integrability can be exploited to construct particularly accurate operator splitting methods to solve the equations of motion of $H$.
However, in many cases, for example in problems related to planetary motion, it is computationally expensive to obtain the exact solution to $H_0$.

In this paper we present a new family of embedded operator splitting (EOS) methods which do not use the exact solution to $H_0$, but rather approximate it with yet another, embedded operator splitting method.
Our new methods have all the desirable properties of classical methods which solve $H_0$ directly. 
But in addition they are very easy to implement and in some cases faster.
When applied to the problem of planetary motion, our EOS methods have error scalings identical to that of the often used Wisdom-Holman method but do not require a Kepler solver, nor any coordinate transformations, or the allocation of memory.
The only two problem specific functions that need to be implemented are the straight-forward kick and drift steps typically used in the standard second order leap-frog method. 
\end{abstract}


\section{Introduction}
One of the main difficulties when calculating the orbital motion of planets over long timescales is the large separation of timescales.
The orbital period of planets around their host star can be as short as a fraction of a day while the age of the system can be billions of years.
Only the development of accurate numerical integration methods and the advent of fast computers have made it possible to predict the orbital evolution of planetary systems over billions of years.

Ignoring planet-planet interactions, the problem of planetary motion is called \textit{integrable}, meaning we can calculate the positions of all planet at any arbitrary moment in the future (or past) almost instantaneously.
Many numerical methods take advantage of the fact that the Hamiltonian system of interest is \textit{near-integrable}, meaning although planet-planet interactions exist, they are small compared to the dominant Keplerian motion. 
One such integrator is the Wisdom-Holman integrator \citep{WisdomHolman1991}, which we will refer to hereafter as WH. 
In addition to the original second order WH integrator, many generalizations and higher order variants have been developed \citep[see][for an overview]{Rein2019b}.
Whereas the WH integrator has become a standard tool for many calculations in astrophysics, the ideas around symplectic integrators for near-integrable systems are much more general and have applications in many different fields \citep[see e.g.][]{Blanes2016,Hairer2006}.

One ingredient of the WH integrator is a \textit{Kepler solver}. 
As the name suggests, it solves Kepler's equation which is in turn required to solve the integrable part of the motion.
Because Kepler's equation is a transcendental equation, it cannot be solved algebraically and typically a series expansions is required.
Although this might lead to philosophical discussions of whether one should really call the Keplerian motion of planets integrable, it is not an issue in practice where all calculations are only performed to some finite precision\footnote{Typically only 16 decimal digits are available for calculations involving numbers represented by the IEEE754 double floating point standard.}.
It has been a long-standing task for mathematicians and astronomers to develop particularly fast, reliable, and un-biased Kepler solvers \citep[the Kepler solver that we will use for comparison tests in this paper is that by][]{ReinTamayo2015}.

In this paper we present a new type of integrator for the near-integrable $N$-body problem of planetary motion which does not require a Kepler solver. 
This dramatically simplifies the algorithms and makes them particularly well suited for situations where complex algorithms are discouraged such as in SIMD (Single Instruction Multiple Data) architectures, GPUs (graphics cards), FPGAs (Field Programmable Gate Arrays), other specialized computing hardware, or when used in conjunction with auto-differentiation methods.

Despite our methods' simplicity, they achieve an efficiency comparable to that of the WH integrator and its higher order extensions.
In certain cases we were even able to measure a speed-up of a factor of 2-3 over the WH integrator. 
Our family of method is very flexible and allows one to construct arbitrarily high order methods.

\section{Hamiltonian Splitting}
We present our new family of methods in the context of the classical $N$-body problem of planetary motion.
However, the methods are equally applicable to many other Hamiltonian systems where one can split the Hamiltonian into one dominant and one perturbation part.

We begin be examining the Hamiltonian $H$, which via Hamilton's equations, governs the evolution of the system.
Suppose we have $N$ gravitationally interacting particles. 
Let us choose the particle with index $i=0$ as the central object around which all other particles orbit on approximately Keplerian orbits. 
The Hamiltonian of this system consists of a kinetic part $T$ plus a potential part $U$,
\begin{eqnarray}
   H=T+U. 
\end{eqnarray}
Let us further split $T$ and $U$ into
\begin{eqnarray}
    T = \sum_{i=0}^{N-1} T_i & \quad \quad\text{and} \quad\quad&
    U = \sum_{i=0}^{N-1}  \sum_{j=i+1}^{N-1} U_{ij},
\end{eqnarray}
where $T_i$ corresponds to the kinetic energy of particle $i$ and the potential $U_{ij}$ corresponds to the interaction potential between particles $i$ and $j$,
\begin{eqnarray}
    T_i = \frac{p_i^2}{2m_i}& \quad \quad\text{and} \quad\quad&
    U_{ij} = -\frac{Gm_im_j}{|r_i-r_j|}.
\end{eqnarray}
Here, $p_i$ and $r_i$ are the canonical momenta and coordinates of particle $i$, and $G$ is the gravitational constant. 

\subsection{TU Splitting}
We can construct one step of a second order splitting method by
\begin{eqnarray}
   \hat U\left(\frac\tau2\right)
   \hat T\left(\tau\right)
   \hat U\left(\frac\tau2\right). \label{eq:leapfrog}
\end{eqnarray}
In the above notation $\hat U\left(\tau/2\right)$ corresponds to an operator advancing a system under the influence of Hamiltonian $U$ forward for a time $\tau/2$. 
The method in Eq.~\ref{eq:leapfrog} is the well known leap-frog integrator, sometimes also referred to as the St\"ormer-Verlet method.
It is easy to show that, since it is a second order method, the relative energy error for the method in Eq.~\ref{eq:leapfrog} scales as 
\begin{eqnarray}
    \frac{\Delta E}{E} \sim \tau^2.
\end{eqnarray}
Both operators for $\hat U$ and $\hat T$ are trivial to implement. 
$\hat U$ is the drift step during which the velocities remain constant and only the positions change.
$\hat T$ is the kick step during which the positions remain constant and only the velocities change.
Let us reiterate the simplicity of this method.
We only need to implement two functions.
One which moves particles along straight lines according to the velocity. 
And one which updates the velocities due to gravitational forces acting on stationary particles.

\subsection{Wisdom-Holman Splitting}
Instead of splitting the Hamiltonian into $T$ and $U$, the standard Wisdom-Holman algorithm, splits the Hamiltonian into
\begin{eqnarray}
    A &=& \sum_{i=0}^{N-1} T_i + \sum_{i=1}^{N-1} U_{i0}\quad \quad\text{and} \label{eq:whA}\\
    B &=& \sum_{i=1}^{N-1}  \sum_{j=i+1}^{N-1} U_{ij} \label{eq:whB}.
\end{eqnarray}
The Hamiltonian $A$ describes the Keplerian motion of particles around the central object. 
On the other hand, the Hamiltonian $B$ describes the interactions between all particles other than the central object. 
Since all particles $i>0$ are on nearly Keplerian orbits, we have $|A|\gg|B|$. 

We can construct one step of a second order operator splitting method (the standard WH integrator) using
\begin{eqnarray}
    \hat A\left(\frac\tau2\right)
    \hat B\left(\tau\right)
    \hat A\left(\frac\tau2\right). \label{eq:wh}
\end{eqnarray}
Because $|B|/|A|\approx \epsilon$ with $\epsilon\ll1$, the relative energy error for the method in Eq.~\ref{eq:wh} scales as 
\begin{eqnarray}
    \frac{\Delta E}{E} \sim \epsilon \tau^2. \label{eq:whscaling}
\end{eqnarray}
This is a factor of $\epsilon$ smaller than the error for the leap-frog method in Eq.~\ref{eq:leapfrog}. 
If we focus on a planetary system with planet masses similar to the giant planets in our own Solar System, then $\epsilon \approx 10^{-3}$. 
Thus, for the same timestep $\tau$ the Wisdom Holman integrator is roughly three orders of magnitude more accurate than the leap frog integrator.
This is a significant improvement in terms of accuracy but comes at the expense of now having to solve the equations of motion for Hamiltonian $A$.
Solving $A$ is non-trivial and involves an iterative approximation to the solution of Kepler's equation. 
There are further subtleties such as the precise coordinate systems to use for the splitting into $A$ and $B$ \citep[see e.g.][]{ReinTamayo2019}.
We here ignore these subtleties but simply point out that they contribute to the complexity of the algorithm, both from a conceptual and implementation point of view.

\subsection{Triple Splitting} \label{sec:newsplitting}
In our new method, we take the Wisdom-Holman splitting from Eqs.~\ref{eq:whA}~and~\ref{eq:whB} and further split $A$ further into two parts 
\begin{eqnarray}
    A_1 = \sum_{i=0}^{N-1} T_i & \quad \quad\text{and} \quad\quad& A_2 = \sum_{i=1}^{N-1} U_{i0} \label{eq:A}.
\end{eqnarray}
The following is then a new second order operator splitting method 
\begin{eqnarray}
  \hat   A_1\left(\frac\tau4\right)
  \hat   A_2\left(\frac\tau2\right)
  \hat   A_1\left(\frac\tau4\right)
  \hat   B\left(\tau\right)
  \hat   A_1\left(\frac\tau4\right)
  \hat   A_2\left(\frac\tau2\right)
  \hat   A_1\left(\frac\tau4\right). \label{eq:hyla}
\end{eqnarray}
One way to think about this method is that it simply approximates the operator $\hat A$ with a leap-frog step using $\hat A_1$ and $\hat A_2$.
Just like the standard leap-frog method (but contrary to the Wisdom-Holman method) all the operators in this method are trivial to implement and only involve (partial) kick and drift steps.
Note that the operators $\hat A_1$ and $\hat A_2$ scale as $O(N)$.
Only the operator $\hat B$ scales as $O(N^2)$. 

Finally, note that we still have an operator splitting method which separates the dominant motion from perturbations, i.e. $|A_1+A_2|\gg|B|$.
Thus we might hope to recover the excellent error scalings of the Wisdom-Holman method but without the need to solve Kepler's equation or perform any coordinate transformations.

\begin{figure*}
    \centering
    \resizebox{0.8\textwidth}{!}{\includegraphics{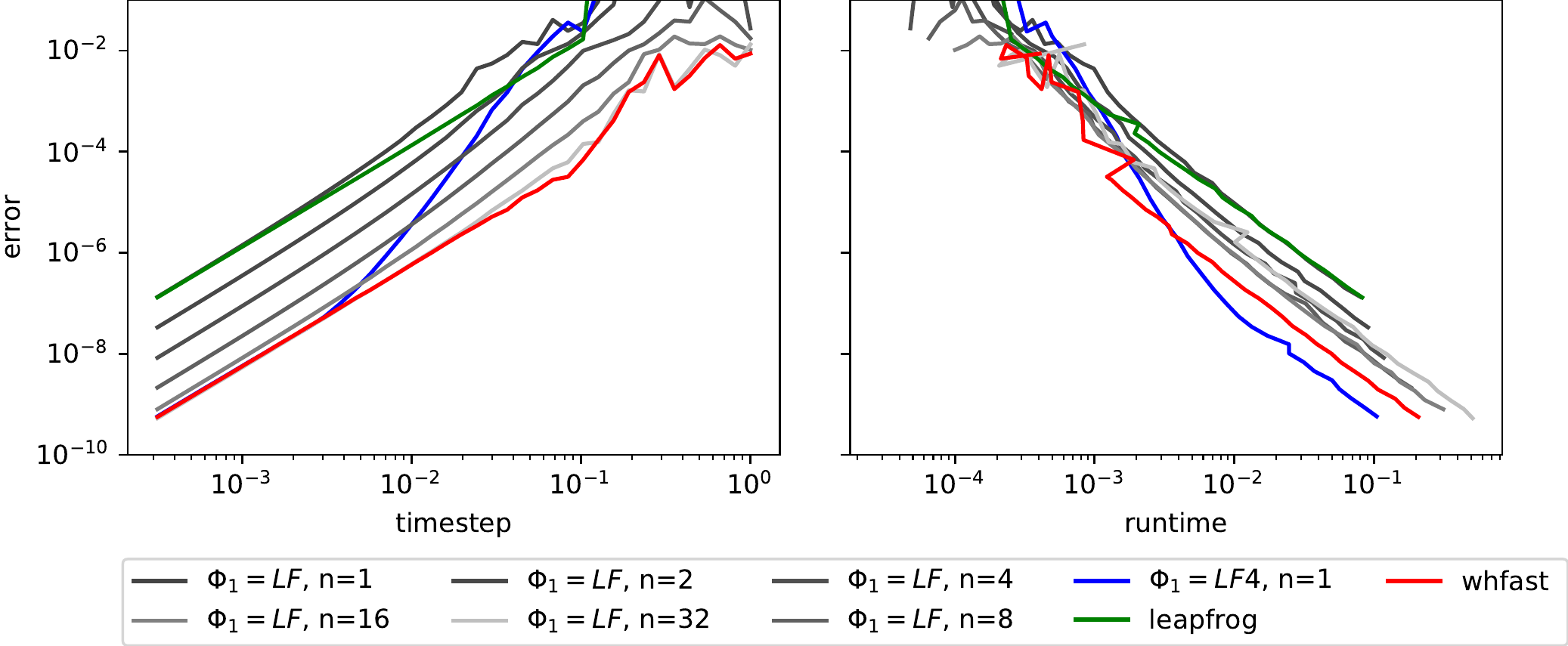}}
    \caption{Relative energy error as a function of the timestep and runtime in a two planet system. The EOS methods shown here use the second order leap-frog method for both $\Phi_0$ and $\Phi_1$.
    In the right hand plot, curves lower and further to the left are more efficient. }
    \label{fig:lf}
\end{figure*}
\begin{figure*}
    \centering
    \resizebox{0.8\textwidth}{!}{\includegraphics{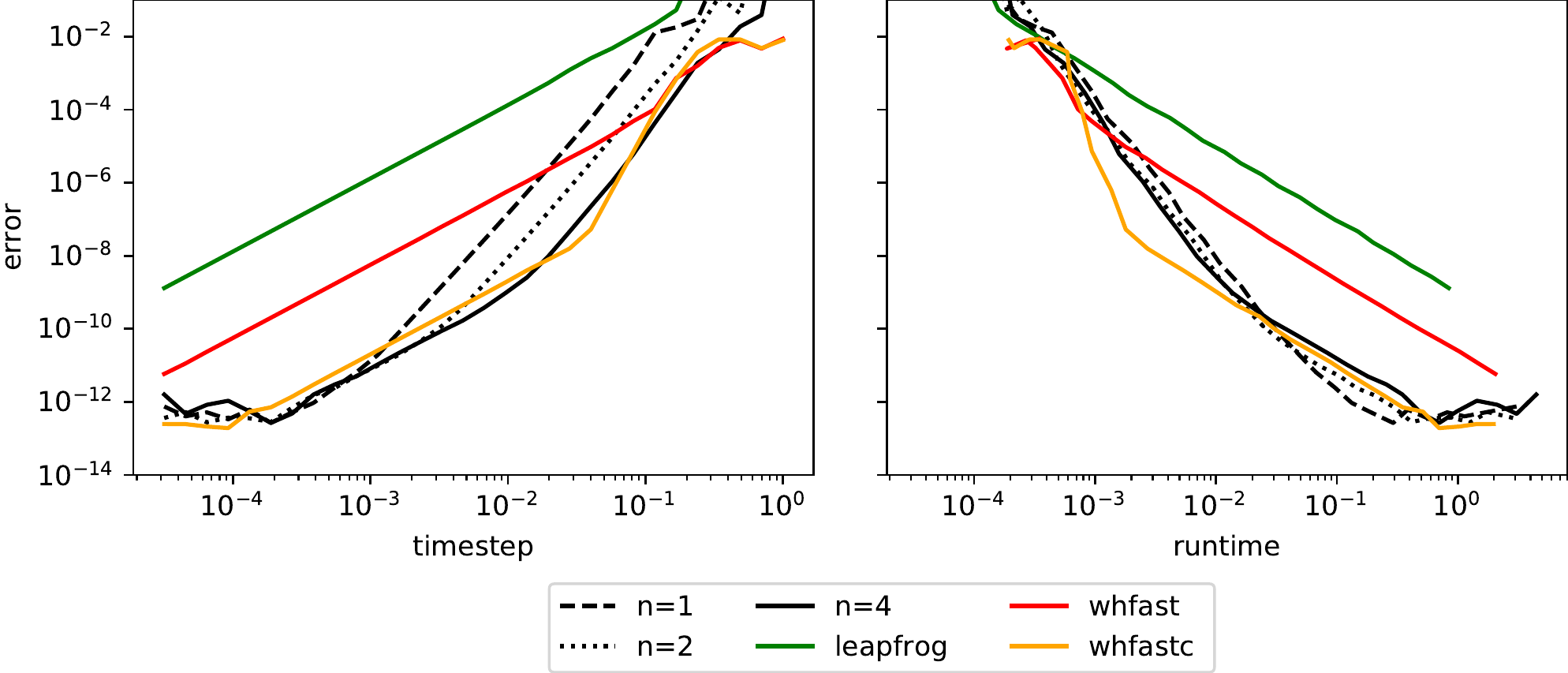}}
    \caption{Relative energy error as a function of the timestep and runtime in a two planet system. The integrators shown here use  $\Phi_0=LF(4,2)$ and $\Phi_1=LF4$ with varying $n$.}
    \label{fig:42}
\end{figure*}

\section{Generalization}
In Section \ref{sec:newsplitting}, we introduced a new splitting method by embedding a second order leap-frog method inside another second order leap-frog method. 
To make this method useful in practice, we will now generalize this to an embedding of an arbitrary operator splitting method $\Phi_1$ within another arbitrary operator splitting method $\Phi_0$. 

The outer operator splitting method $\Phi_0$ needs to invoke two sub-routines. 
One of them is the routine which solves the interaction part of the Hamiltonian, $B$.
The other is the inner operator splitting method $\Phi_1$ which will provide an approximation to the solution of the Keplerian part of the Hamiltonian, $A$. 
Any arbitrary composition method can be used for $\Phi_0$. 
For example, we can choose methods which use a pre- and post-processor\footnote{These are often referred to as \textit{symplectic correctors}, see \cite{Wisdom1996}}. 
We can also choose methods with arbitrarily high order. 
Further, note that since the Hamiltonian $B$ is a potential, we can also calculate the derivatives of the accelerations\footnote{The WH method can also make use of this fact. See the idea of using a \textit{modified kernel} in \cite{Wisdom1996}.}.
Operator splitting methods which make use of this can be particularly efficient \citep{Chin1997,Blanes2016}.

We not only have complete freedom when it comes to choosing an operator splitting method $\Phi_0$ but also when it comes to choosing $\Phi_1$. 
As before, we can choose methods of arbitrary order, with or without pre- and post-processors.
And since $A_2$ is again a potential, we can also use methods which use the derivatives of accelerations.

We have argued that $\Phi_1$ merely provides an approximation of the otherwise difficult to calculate operator $\hat A$ required by $\Phi_0$. 
To improve the approximation we can either choose a high order method for $\Phi_1$, or simply use a low order method and reduce the timestep. 
Say, the outer method $\Phi_0$ requires an approximation for $\hat A(\tau/2)$ as in Eq.~{\ref{eq:hyla}.
Then, instead of just taking one timestep of length $\tau/2$ with $\Phi_1$, we can also take $n$ timesteps with $\Phi_1$, each advancing the solution by $\tau/(2n)$. 
In all symmetric integrators, we can combine the first and last steps at the beginning and end of consecutive timesteps. 
For example, in the case of the WH integrator, the two $\hat A(\tau/2)$ operators can be combined into one $\hat A(\tau)$ operator.
We thus use the definition of $n=1$ corresponding to replacing $\hat A(\tau)$ with one step of $\Phi_1$, $n=2$ with two steps of $\Phi_1$ (one for each  $\hat A(\tau/2)$), and so forth. 

The results which we present below focus on drift-kick-drift type integrators, i.e. integrators which start with a drift step. 
We do not expect any significant improvement to either performance or accuracy from using kick-drift-kick type integrators.

In summary, to completely characterize a method in our new family, we need to specify the outer operator splitting method $\Phi_0$, the inner operator splitting method $\Phi_1$, and the number of sub-steps $n$.  

\section{Results}
In this section we present simulation results which test the accuracy and efficiency of our new family of methods. 
We use a system of units where $G=1$.
All simulations include a stellar object of mass $m_0=1$ and two lower mass objects with masses $m_1=m_2=10^{-3}m_0$.
The smaller objects (planets) are initially orbiting the central object (the star) with semi-major axes $a_1=1$, $a_2=1.6$ and eccentricities $e_1=e_2=0.1$.
In all plots, we use the orbital period of the inner planet as the unit of time.

We will choose the methods for $\Phi_0$ and $\Phi_1$ from the following list of symplectic operator splitting methods. 
This is by no means a comprehensive list of possible methods we could try. 
We here focus on the most promising lowest order methods, and a few high order methods for illustration purposes. 
\begin{itemize}[leftmargin=10pt,labelindent=0pt,itemindent=-10pt,align=left]
    \item $LF$: the standard second order leap-frog or St\"ormer-Verlet method.
    \item $LF4$: A fourth order Suzuki-Yoshida method using three force evaluations per timestep \citep{Creutz1989}.
    \item $LF8$: An eighth order method with 17 function evaluations per timestep \citep{Mclachlan1995b}.
    \item $LF(4,2)$: A second order method using two function evaluations per timestep \citep{Mclachlan1995}. This method has generalized order~$(4,2)$, i.e. the dominant error term for small timesteps is $O(\epsilon \tau^4 + \epsilon^2 \tau^2)$ and there is no error term $O(\epsilon \tau^2)$. 
    \item $LF(8,6,4)$: A fourth order method with generalized order (8,6,4) using seven function evaluations per timestep \citep{Blanes2013}. 
        When used with a WH-type splitting and a Kepler solver, it is referred to as $SABA(8,6,4)$. The dominant error term is $O(\epsilon \tau^8 + \epsilon^2 \tau^6 + \epsilon^3 \tau^4)$.
    \item $PMLF4$: A fourth order method with only one modified force evaluation per timestep \citep{Blanes1999}. This method also includes pre- and post-processing stages. 
\end{itemize} Just as the basic leap-frog algorithm, all methods above internally only use one problem-specific drift and one (modified) kick function.
They only differ in how often they call these functions, in which order, and for what length.
The methods $LF(4,2)$, $LF(8,6,4)$ above are designed for near-integrable (or perturbed) systems.

\begin{figure*}
    \centering
    \resizebox{0.8\textwidth}{!}{\includegraphics{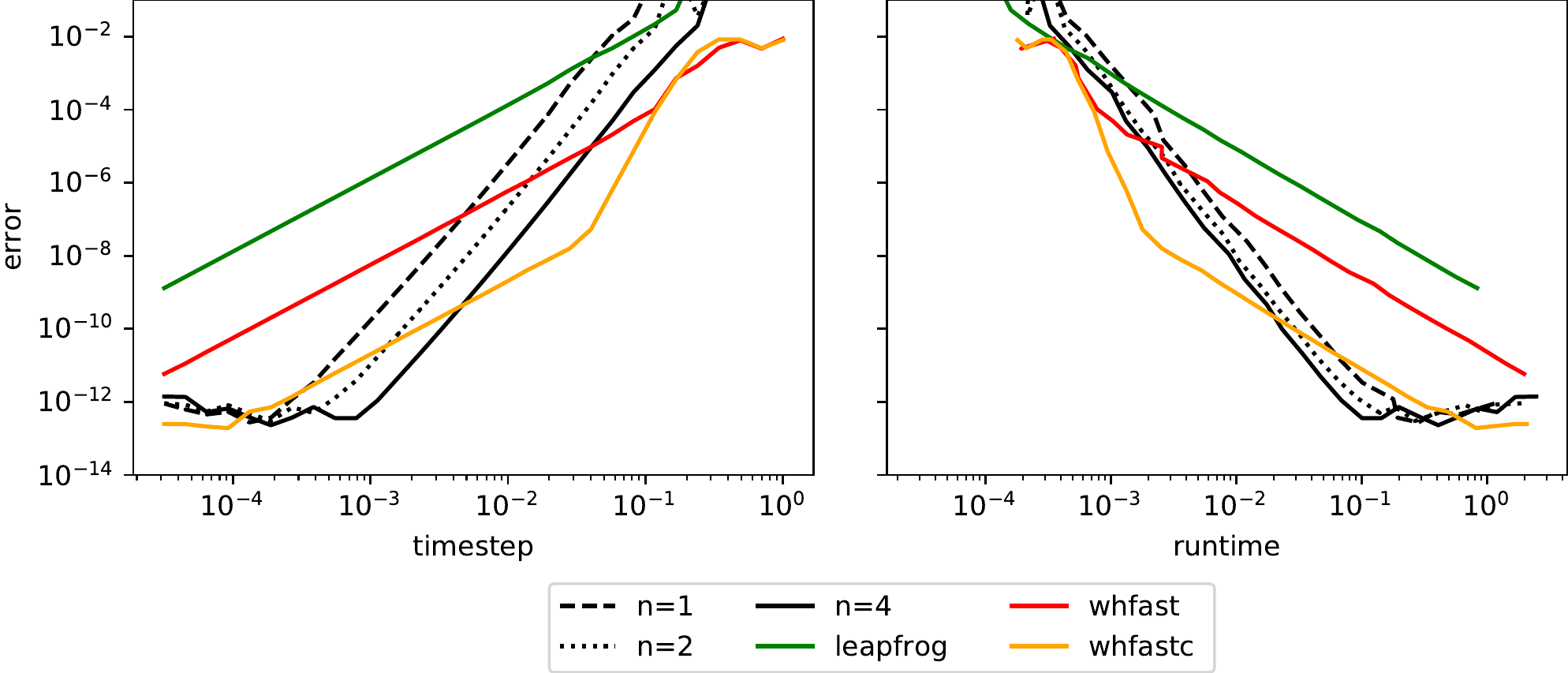}}
    \caption{Relative energy error as a function of the timestep and runtime in a two planet system. The integrators shown here use $\Phi_0=PMLF4$ and $\Phi_1=LF4$ with varying $n$.}
    \label{fig:4}
\end{figure*}
\begin{figure*}
    \centering
    \resizebox{0.8\textwidth}{!}{\includegraphics{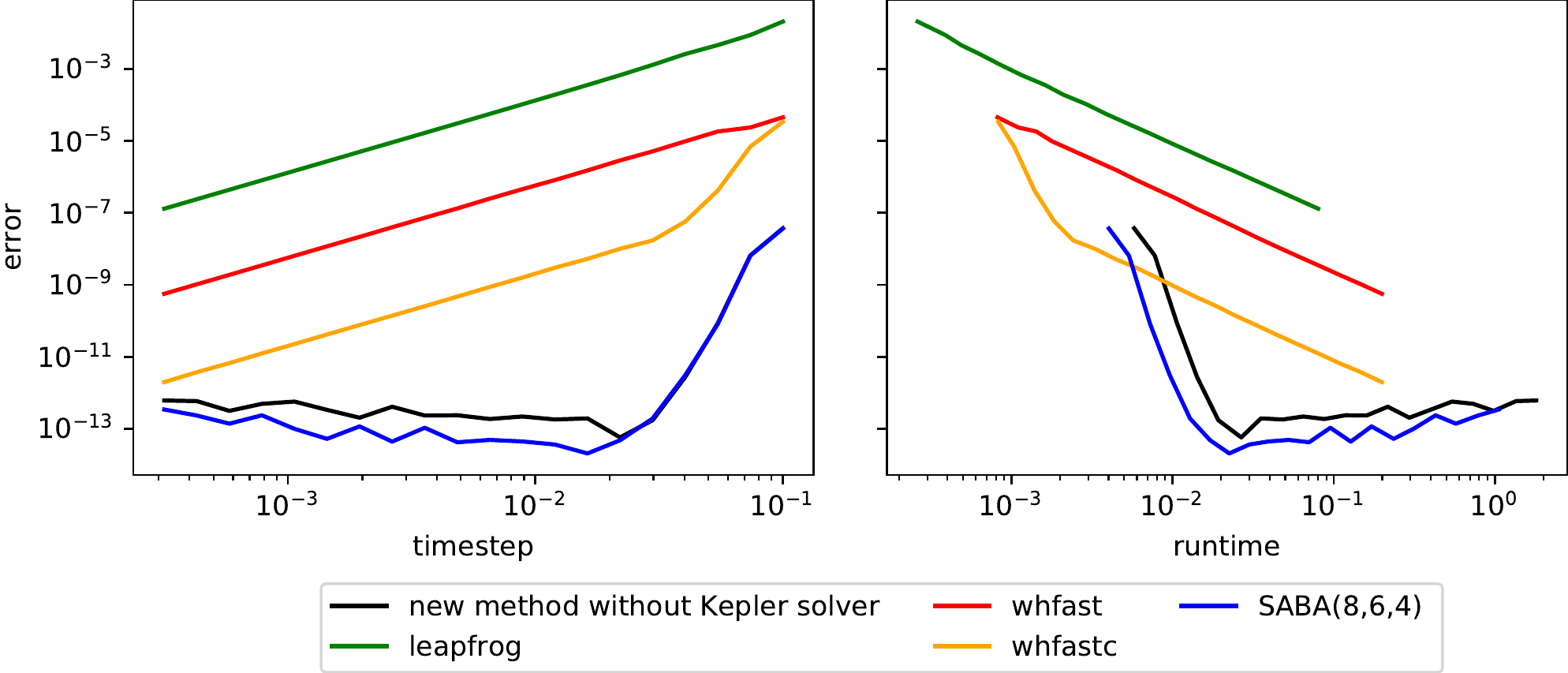}}
    \caption{Relative energy error as a function of the timestep and runtime in a two planet system. The integrator shown here uses $\Phi_0=LF(8,6,4)$, $\Phi_1=LF8$, and $n=1$.}
    \label{fig:764}
\end{figure*}
\subsection{Embedded leap-frog}
We start with results of simulation using $\Phi_0=\Phi_1=LF$ and varying $n$.
Note that we recover the method in Eq.~\ref{eq:hyla} for $n=2$.
In Figure~\ref{fig:lf} we plot the relative global energy error as a function of the timestep on the left panel, and the relative global energy error as a function of the runtime on the right panel.
We also over-plot the results of the standard leap-frog integrator and WHFast, an implementation of the WH integrator \citep{ReinTamayo2015}.
In all cases, we measure the maximum energy error over approximately $160$~periods of the inner planet.
The runtime is measured in seconds and all simulations were performed on a 3.3 GHz Intel Core i7 processor.

We can see that if we replace the operator $\hat A$ with just one single leap-frog step, we have gained nothing over the basic leap-frog method from Eq.~\ref{eq:leapfrog}. 
However, if we keep on improving our approximation of $\hat A$ by increasing $n$, then we eventually approach the accuracy of the WH integrator at around $n\sim32$. 
Having $n=32$ sub-steps may seem like a lot, but note that solving $\hat A$ is $O(N)$.
Depending on the specific problem, this might not be the slowest part of the algorithm when compared to solving $\hat B$, with is $O(N^2)$. 
As one can see on the right panel, we can find a value $n$ for which the new method is only marginally less efficient than the WH method at the same accuracy.

This examples shows that even the most basic member in our new family of methods can indeed reproduce the advantageous scaling properties of the WH method in near-integrable systems.

Instead of using a second order method with $n=32$ steps to provide an approximation of $\hat A$, we can also choose a higher order method to do the same with fewer steps $n$.
In Figure~\ref{fig:lf} we also plot the results for $\Phi_0=LF$, $\Phi_1=LF4$, and $n=1$. 
We can see that a single timestep with a fourth order method is enough to achieve the same accuracy as the WH integrator for timesteps smaller than $10^{-2}$. 
Most importantly, we can see in the right panel that this new method is about a factor of two faster than the WH method at accuracies better than about $10^{-5}$.

\subsection{Generalized order (4,2)}
In this section we present the result for a new integrator with the same scalings as a WH integrator with symplectic correctors. 
The WH method with correctors has a generalized order of (k,2), meaning the leading error term is $O(\epsilon \tau^k + \epsilon^2 \tau^2)$, a factor of $\epsilon$ smaller than for the WH integrator without correctors for small timesteps.
In principle $k$ can be arbitrarily large, but typically a value of $k\leq 17$ is sufficient.

To illustrate this, we set $\Phi_0=LF(4,2)$ and $\Phi_1=LF4$. We vary $n$ from 1 to~4.
Figure~\ref{fig:42} shows that the integrator is indeed approaching the same scaling as the WH integrator with correctors (shown as WHFastC in the plot). 
For large timesteps, we need $n$ to be 2 or 4 to achieve the same accuracy as WHFastC.
For small timesteps, we approach the same accuracy even with $n=1$. 
The efficiency of our integrator with $n=1$ exceeds that of WFastC by a factor of about $2-3$ for small timesteps. 
This efficiency gain can be easily understood: we replace the complicated Kepler-solver with a single leap-frog step, but achieve the same accuracy.
Note that for very small timesteps the integrators approach machine precision.

\subsection{Fourth order}
Here we construct a fourth order method by choosing fourth order methods for both $\Phi_0$ and $\Phi_1$. Specifically, we set $\Phi_0=PMLF4$ and $\Phi_1=LF4$.

The results are shown in Figure~\ref{fig:4}. 
One can clearly see that the methods have indeed a scaling of $O(\tau^4)$ until they reach machine precision.
For sufficiently small timesteps, unsurprisingly, the fourth order methods are more efficient than the second order WH method.
This is a slightly unfair comparison as higher order versions of the WH integrator do exist \citep[see below and also][for a recent review]{Rein2019b} but illustrates that we can easily construct methods of arbitrary order.

\begin{figure*}
    \centering
    \resizebox{0.8\textwidth}{!}{\includegraphics{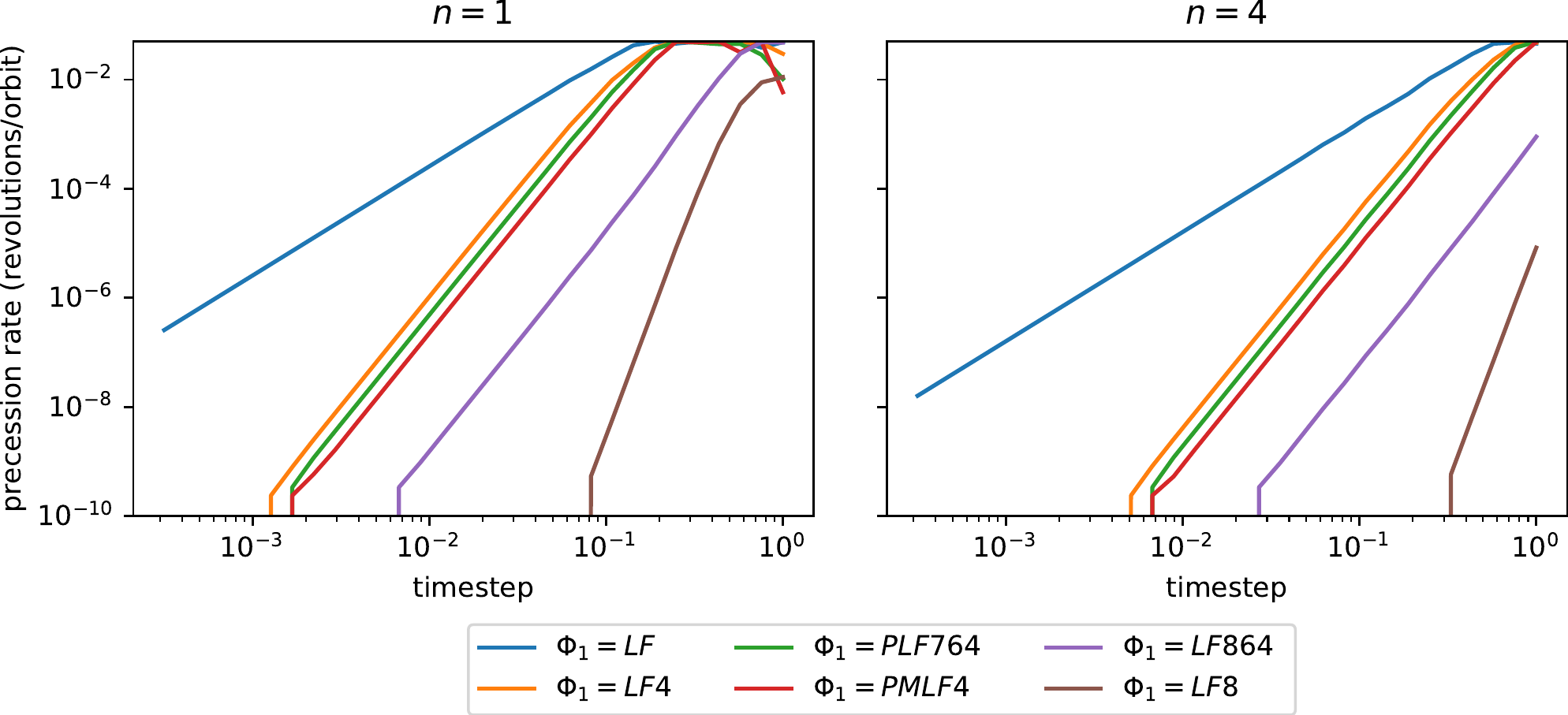}}
    \caption{Artificial precession rate as a function of the timestep in a one planet system. The EOS methods shown here use the second order leap-frog method for $\Phi_0$. The methods in the left panel use  $n=1$, those in the right panel use $n=4$. The WH method is not shown here as it has not artificial precession.}
    \label{fig:prec}
\end{figure*}

\subsection{High generalized order (8,6,4)}
Let us now attempt to construct a very high order integrator.
For this we use $\Phi_0=LF(8,6,4)$, $\Phi_1=LF8$, and $n=1$. 
The results are shown in Figure~\ref{fig:764}. 

This new method now converges to machine precision for timesteps smaller than $5\%$ of the orbital period of the inner most planet.
For the range of timesteps typically of interest, the error in the new method scales as $\tau^8$, although formally it is only a fourth order method for extremely small $\tau$.
We also over-plot the curves corresponding to $SABA(8,6,4)$.
The only difference between our method and $SABA(8,6,4)$ is that $SABA(8,6,4)$ uses a Kepler solver \citep{Blanes2013}, whereas we use $\Phi_1$.

The efficiency of $SABA(8,6,4)$ is slightly better than that of our new method by about a factor of 1.5.
The reason why we cannot beat a very high order method with a Kepler solver is that the timesteps are very large, between $5\%$ and $10\%$ of the orbital period.
Thus to resolve the Keplerian motion accurately (note that we are achieving accuracies better than $10^{-9}$ for timesteps as large as $10\%$), we either need a high order method for $\Phi_1$ (we have chosen an 8th order method), or a lower order method with a large number of steps $n$. 
For a Kepler solver on the other hand, the time to solve Kepler's equation does not depend on the timestep.
Further note that the round-off error is slightly worse for our method than for SABA(8,6,4). 
This is because round-off error accumulated in the 17 stages of $\Phi_1$, whereas less round-off error occurs in the Kepler-solver.
Nevertheless, note that getting within a factor of 1.5 of the performance of SABA(8,6,4) but without the need to implement a Kepler solver is remarkable.

\subsection{Artificial precession due to $\Phi_1$}
One disadvantage of the EOS methods is the existence of artificial precession even in a one planet system.
The WH method does not exhibit any artificial precession for a one planet system because it solves the Kepler problem exactly.
However, the WH method does of course exhibit artificial precession when there are multiple planets \citep[see][for a discussion on how this affects secular frequencies]{Rein2019c}. 

To illustrate this issue and help with the choice of $\Phi_1$ we measure the artificial precession rate for a one planet system with $e=0.1$ in units of revolutions per orbit.
For comparison, the precession rate of Mercury is about $5.75$~arc-seconds per year or about $10^{-6}$ revolutions per orbit. 
The results are shown in Fig.~\ref{fig:prec} for methods with varying $\Phi_1$ and $n$.
Note that the choice of $\Phi_0$ does not affect the precession rate\footnote{The choice of $\Phi_0$ does slightly affect the precession rate because the longest timestep taken by $\Phi_1$ depends on $\Phi_0$.}.
All results shown use $\Phi_0=LF$.

One can see that the precession rate is proportional to the second power of the timestep when a second order method is used for $\Phi_1$, proportional to the fourth power when a fourth order method is used, and so on. 
Note that because we are not integrating a perturbed system ($A_1$ and $A_2$ are approximately equal in magnitude) only the order of the integrator matters, not the generalized order.
Increasing $n$ has the same effect as reducing the timestep. 

To accurately capture the dynamical evolution of a planetary system, the artificial precession rate needs to be slower than any precession rate due to planet-planet interactions. 
As an example, we can consider Mercury in the Solar System and the EOS method with $n=4$ and $\Phi_1=LF4$.
We start resolving the physically relevant precession rate if the timestep is smaller than about 20 steps per orbit.
Note that the timescale of periastron passage is shorter for orbits with higher eccentricities, and thus the artificial precession will be larger for very eccentric orbits.
We come back to this issue in the discussion section.

\subsection{Suggestions for choosing $\Phi_0$ and $\Phi_1$}
The choice of $\Phi_0$, $\Phi_1$ and $n$ to obtain an optimal performance is problem specific and might require some experimentation.
In the planetary motion case for example, the optimal choice will depend on the number of planets, the planet masses, the typical eccentricity of planets, and the desired accuracy.
But we can provide some general suggestions.

For example, the function evaluations in $\hat B$ are expensive, thus $\Phi_0$ should have as few as possible. 
It therefore makes sense to choose a method optimized for near-integrable systems with high general order if $\epsilon$ is small.
On the other hand, function evaluations are not so much an issue for $\Phi_1$ as evaluating the potential $A_2$ scales as $O(N)$, thus $n$ can be large if $N$ is large. 
Pre- and post-processors are more suited for $\Phi_0$ than $\Phi_1$ because the $\Phi_1$ processors need to get called whenever the result is returned to $\Phi_0$.

\section{Discussion}
In this paper we have introduced a very flexible new family of integrators for perturbed Hamiltonian systems where the dominant part of the Hamiltonian is computationally expensive to solve exactly. 
We achieve this by embedding one operator splitting method within another.
We refer to our family of methods as Embedded Operator Splitting (EOS) methods.

The performance of our EOS methods is comparable to that of standard Wisdom-Holman integrators.
Depending on the precise setup, for example in simulations which do not require extremely high accuracy, a speed-up by a factor of 2-3 can be achieved.

Whereas this speed-up should be useful for many applications, the main benefit of our new methods is how easy they are to implement. 
All that is required are (partial) drift and kick steps. 
This significantly reduced the complexity typically associated with the standard Wisdom-Holman integrator. 
Specifically, there is no need to implement an accurate Kepler solver, Stumpf functions, or any coordinate transformations.
Furthermore, none of the new methods presented here require any memory allocation. They can simply act directly on the particles' positions and velocities. 
All these simplifications make EOS methods particularly well suited for use on graphic cards (GPUs) or other computing architectures with reduced instruction sets and parallel execution models\footnote{It can be impossible for a compiler to predict a branch's execution time if an iteration depends on some parameter only available at run-time (this is the case in many implementations of a Kepler solver). In a SIMD model different threads have to stay synchronized which can lead to a significant overhead.}. 

A further advantage of the EOS methods is that particles always move on straight lines during drift steps ($A_1$). 
This significantly simplifies collision detection algorithms.
Furthermore it is trivial to add, remove, or merge particles during any part of the integrator (sub-)steps without the need to worry about coordinate systems.
These properties suggest that hybrid integration methods which can resolve close encounters and are based on EOS might have particularly desirable properties (Rein et al., in prep). 

Yet another advantage is that our methods can be easily applied to integrate variational equations, or tangent maps \cite[][]{ReinTamayo2016}.
Variational equations in Hamiltonian systems can be used to calculate chaos indicators.
They further play an important role to provide accurate derivatives which are required for gradient-based optimization methods.
Among many other application, such methods are useful when fitting observed transit timing variations (TTVs) of extrasolar planets. 
Once again, the simplicity of our methods, compared to the complexity of tangent maps of the Keplerian motion \citep{MikkolaInnanen1999}, make these tasks much more feasible. 
Also note that the presence of iterations and conditional break-out conditions in several parts of the WH method make the usage of auto-differentiation algorithms more challenging.  
Auto-differentiation methods are straightforward to use with EOS methods.

Like all integrators, the EOS methods fail when the timestep is too large to resolve the shortest timescale in the problem. 
In problems of planetary motion, the shortest timescale is often a planet's perihelion passage timescale. 
This timescale can be significantly shorter than the orbital period for high eccentricities.
One might be tempted to argue that the failure of the EOS methods in those cases is due to the inaccurate approximation of the Keplerian motion given by $\Phi_1$. 
However, such an argument would be misleading. 
Even if $\Phi_1$ were exact, the EOS method now just being a WH method, the timestep would still need to be small enough to resolve the perihelion passage \citep{Wisdom2015}.
A simple argument for this is that an integrator cannot possibly give physical results if it does not resolve the shortest timescale in the problem.
In such a case, neither EOS nor WH methods can be trusted to reproduce the correct dynamics without reducing the timestep. 
However, the failure of the EOS integrators will be more noticeable on short timescales in the form of large energy errors.
This might be helpful for alerting a user to a potential problem.

The discussion in this paper has focused on planetary motion. 
However, the concepts introduced here can be used in many other areas.
Any Hamiltonian system which can be split into a dominant and a perturbation part can use our embedded operator splitting methods.
In particular, the methods can be used in cases where the dominant part is not integrable at all.
Two examples are non-Keplerian potentials in galactic dynamics or strong gravity regimes. 
A somewhat similar algorithm was recently proposed for solve the semiclassical time-dependent Schr\"odinger equation by \cite{Blanes2020}.

All integrators presented in this paper are evidently symplectic. 
They are simply compositions of symplectic methods. 
In fact, one can make the theoretical argument that the methods presented here are formally symplectic, whereas standard Wisdom-Holman methods are not, because they contain a series expansion in the Kepler solver which must be truncated at some point. 
Although practically this makes no difference as one rapidly converges to machine precision.

All methods presented in this paper have been implemented in the REBOUND integrator package available at \url{https://github.com/hannorein/rebound}. 
However, given how easy it is to implement our new methods, we encourage the reader to give it a try themselves!

\section*{Acknowledgments}
I would like to thank Samuel Hadden for an in-depth review which helped to significantly improve this manuscript.
I would like to further thank Daniel Tamayo and Scott Tremaine for proofreading an early version of this manuscript and many helpful discussions on symplectic integrators over the years.
I am also grateful to Sergio Blanes for many discussions and recommendations regarding geometric integration methods.
This research has been supported by the NSERC Discovery Grant RGPIN-2014-04553 and the Centre for Planetary Sciences at the University of Toronto Scarborough.
This research was made possible by the open-source projects \texttt{REBOUND} \citep{ReinLiu2012}, 
\texttt{Jupyter} \citep{jupyter}, \texttt{iPython} \citep{ipython}, 
and \texttt{matplotlib} \citep{matplotlib, matplotlib2}.

\bibliography{full}

\label{lastpage}
\end{document}

%% file: mnras_jdf.tex
\def \aap{A\&A}

\def \apjl{ApJ}
\def \apjs{ApJS}